\documentclass{iaus}
\usepackage{graphicx}
\usepackage{amssymb}
\usepackage{amsmath}
\usepackage{natbib}
\usepackage{subfigure}
\usepackage{booktabs}

\title[Effects of Magnetic Braking and Tidal Friction on Hot Jupiters] 
{Effects of Magnetic Braking and Tidal Friction on Hot Jupiters}

\author[A.J. Barker \& G.I. Ogilvie]   
{Adrian J. Barker, 
 Gordon I. Ogilvie}

\affiliation{Department of Applied Mathematics and Theoretical Physics, University of Cambridge, Centre for Mathematical Sciences, Wilberforce Road, Cambridge CB3 0WA, UK \break email: ajb268@cam.ac.uk}

\pubyear{2009}
\volume{259}  
\pagerange{100--100}
\date{Dec. 10th, 2008 and in revised form ??}
\jname{Cosmic Magnetic Fields: From Planets, to Stars and Galaxies}
\editors{K.G. Strassmeier, A.G. Kosovichev \& J.E. Beckman, eds.}
\begin{document}

\maketitle

\begin{abstract}
Tidal friction is thought to be important in determining the long-term 
spin-orbit evolution of short-period extrasolar planetary systems. 
Using a simple model of the orbit-averaged effects of tidal friction 
\citep{Eggleton1998}, we analyse 
the effects of the inclusion of stellar magnetic
braking on the evolution of such systems. A phase-plane analysis of a
simplified system of equations, including only the stellar tide
together with a model of the braking torque proposed by
\cite{VerbuntZwaan1981}, is presented. The inclusion of stellar
magnetic braking is found to be extremely important in determining the secular
evolution of such systems, and its neglect results in a very different
orbital history. We then show the 
results of numerical integrations of the full tidal evolution
equations, using the misaligned spin and orbit of the XO-3 system as an example, to study the accuracy
of simple timescale estimates of tidal evolution. We find that it is
essential to consider coupled evolution of the orbit and the stellar
spin in order to model the behaviour accurately. In addition, we
find that for typical Hot Jupiters the stellar spin-orbit alignment 
timescale is of the same order as the
inspiral time, which tells us that if a planet is observed to be aligned,
then it probably formed coplanar. This reinforces the importance of
Rossiter-McLaughlin effect observations in determining the degree of
spin-orbit alignment in transiting systems.
\end{abstract}

\firstsection 
\section{Introduction}

Since the discovery of the first extrasolar planet around a solar-type
star \citep{MQ1995}, observers have now detected 322\footnote{As of
  the day of this session, 4th Nov. 2008 - see http://exoplanet.eu/} planets
around stars outside the solar system. Many of these planets have
roughly Jovian masses and orbit their host stars in orbits with semi-major
axes less than $0.2$ AU, the so-called ``Hot Jupiters'' (HJs). In both
of the giant planet formation scenarios,
core accretion and gravitational instability, we are unable to produce
HJs in situ. We require the formation of close-in
planets much further out ($a \sim $
several AU) in the protoplanetary disc, before a migratory process that brings the planet in
towards the star and to its present location \citep{LBR1996}. 

The formation of systems with giant planets can be thought of as
occurring in two oversimplified stages
\citep{JT2008}. 
During stage 1 the cores of the giant planets are formed, they accrete
gas and undergo migration, driven by the dynamical interaction between the
planets and the gaseous protoplanetary disc (see \citealt{PapProt2007} for a
recent review).  This stage lasts a few Myr until the gas dissipates, by which
time a population of gas giants may exist.  If these form sufficiently
closely packed then stage 2 follows.
This stage lasts from when the disc has dissipated and continues until the
present, and primarily involves gravitational interactions and
collisions between the planets. Recent studies into stage 2
(\citealt{JT2008}; \citealt{CFMR2008}) have shown that this is a chaotic era, in
which planet-planet scatterings force the
ejection of all but a few ($\sim 2-3$) planets from the system in a
period of large-scale dynamical instability lasting $\leq
10^{8}$yr. This mechanism can excite the eccentricities of the planets
to levels required to explain observations. 

Planet-planet scatterings tend also to excite the inclinations of the planets with respect to the initial
symmetry plane of the system, potentially leading to observable
consequences via the Rossiter-McLaughlin (RM) effect, which allows a
measurement of $\lambda$, the sky-projected angle of
misalignment of the stellar spin and the orbit \citep{Winn2008}. 
Misaligned orbits are not predicted from stage 1
alone, so if $\lambda$ is measured
to be appreciably nonzero in enough systems, then it could be seen as
evidence for planet-planet
scattering, or alternatively Kozai migration (see
\citealt{Fabrycky2007}). This is because 
gas-disc migration does not seem able to excite orbital inclination
(\citealt{LubOg2001}; \citealt{Cresswell2007}). Alternatively, if observed planets are all found with
$\lambda$ consistent with zero, this could rule out planet-planet scattering
or Kozai migration as being
of any importance. One important consideration is that at such close
proximity to their parent stars, strong tidal interactions are 
expected to change $\lambda$ (actually the true spin-orbit
misalignment angle $i = \arccos (\mathbf{\hat
  \Omega \cdot \hat h})$) over time. If tides can significantly change
$\lambda$ since the time of formation, we may have difficulty in
distinguishing between the possible HJ formation mechanisms of
planet-planet scattering, Kozai migration and gas-disc migration. 
Below we approach the problem of studying the effects of
tidal friction on such inclined orbits.

\vspace{-0.38cm}
\section{Model of tidal friction \& magnetic braking}

The efficiency of tidal dissipation in a body is usually parametrised by a
dimensionless quality factor $Q$, defined by:
\[ Q = 2\pi E_{0} \left(\oint-\dot E dt\right)^{-1}, \]
where $E_{0}$ is the maximum energy stored in an oscillation and the
integral represents the energy dissipated over one oscillation period. We find it
convenient to define $Q^{\prime} = \frac{3Q}{2k_{2}}$,
where $k_{2}$ is the second-order potential Love number, since this combination
appears together in the theory. 

$Q^{\prime}$ 
is in principle a function of the tidal forcing frequency and the 
amplitude of the tidal disturbance, and is a
result of complex dissipative processes in each body
\citep{Zahn2008}. Recent work has shown that $Q^{\prime}$ for
solar-type stars depends in a highly
erratic way on the tidal forcing frequency \citep{Gio2007}. In light
of the uncertainties involved in calculating $Q^{\prime}$, and the difficulty
of calculating the evolution when $Q^{\prime}$ depends on tidal
frequency, we adopt a simplified model based on an equilibrium tide with constant lag time,
using the formulation of \citealt{Eggleton1998}. This formulation is
beneficial because it can treat arbitrary orbital eccentricities
and stellar and planetary obliquities. We refer the reader to
\cite{AJB2009} or \cite{ML2002} for the full set of 
equations used.

Observations of solar-type stars show that the mean 
stellar rotational velocity decreases with main-sequence age
\citep{Skumanich1972}, following the relation $\Omega \propto t^{-1/2}$.
Magnetic braking by a magnetised outflowing wind has long been
recognised as an important mechanism for the removal of angular
momentum from rotating stars \citep{WD1967}, and such a mechanism
seems able to explain (on average) most of the
observed stellar spin-down \citep{Barnes2003}.

Here we include the effects of magnetic braking in
the tidal evolution equations, through the inclusion of the magnetic
braking torque of \cite{VerbuntZwaan1981}. This gives the following term into of the equation for
$\mathbf{\dot \Omega}$ \citep{DDLM2002}:
\[ \mathbf{\dot \omega}_{mb} =
- \alpha_{mb} \; \Omega^{2} \; \mathbf{\Omega} \]
where $\alpha_{mb} = 1.5 \times 10^{-14}$ yr, giving a
braking timescale $\sim 2\times 10^{11}$ yr for the Sun.

\vspace{-0.38cm}
\section{Analysis of the effects of magnetic braking on tidal evolution for a
  simplified system}

We study the effects of magnetic braking on a simplified system of a
circular, coplanar orbit under the influence of only the stellar tide and magnetic braking. The following set of
dimensionless equations can be derived \citep{AJB2009}:
\begin{eqnarray}
\label{eqn:ODE1}
\frac{d\tilde \Omega}{d\tilde t} &=&
\tilde n^{4}\left(1-\frac{\tilde \Omega}{\tilde n}\right) -
A \: \tilde \Omega^{3} \\
\frac{d\tilde n}{d\tilde t} &=& 3 \: \tilde n^{\frac{16}{3}}\left(1-\frac{\tilde \Omega}{\tilde n}\right),
\label{eqn:ODE2}
\end{eqnarray}
where we have normalised the stellar rotation $\Omega$ and orbital
mean motion $n$ to the orbital frequency at the
stellar surface, together with a factor $C$ (to some
power). $C$ is the ratio of the orbital angular momentum of an orbit with semi-major
axis equal to the stellar radius $R_{\star}$, of a mass $m_{p}$, with
the spin angular momentum of an equally rapidly rotating star of radius
$R_{\star}$, mass $m_{\star}$ and dimensionless radius of gyration
$r_{g}$. The reduced mass is $\mu=\frac{m_{\star}m{p}}{m_{\star}+m_{p}}$. $C$ is important for
classifying the stability of the equilibrium curve $\tilde \Omega = \tilde
n$ in the absence of magnetic braking, and it can be shown that this
equilibrium is stable if $\tilde n \leq 3^{-\frac{3}{4}}$ i.e. no more than a quarter of the total angular
momentum can be in the form of spin angular momentum \citep{Hut1980}. We have thus defined the
following dimensionless quantities:
\vspace{-0.2cm}
\begin{eqnarray*}
  \tilde \Omega &=& \Omega\left(\frac{R_{\star}^{3}}{G(m_{\star}+m_{p})}
  \right)^{\frac{1}{2}}C^{-\frac{3}{4}},
  \;\;\;\;\; \tilde n = n\left(\frac{R_{\star}^{3}}{G(m_{\star}+m_{p})}
  \right)^{\frac{1}{2}}C^{-\frac{3}{4}}, \;\;\;\;
  C = \frac{\mu R_{\star}^{2}}{I_{\star}} = \frac{\mu}{r_{g}^{2} \: m_{\star}}, \\
  \tilde t &=& \sqrt{\frac{G(m_{\star}+m_{p})}{R_{\star}^{3}}} \: 
    \left(\frac{9}{Q^{\prime}}\right)\left(\frac{m_{p}}{m_{\star}}\right) 
    \: C^{\frac{13}{4}} \; t, \\
    A &=& \alpha_{mb} \:
    \: \sqrt{\frac{G(m_{\star}+m_{p})}{R_{\star}^{3}}} \:
    \left(\frac{Q^{\prime}}{9}\right)\left(\frac{m_{\star}}{m_{p}}\right)
    \left(\frac{\mu R_{\star}^{2}}{I_{\star}}\right)^{-\frac{7}{4}} 
    \simeq 100 \:
    \left(\frac{Q^{\prime}}{10^{6}}\right).
\end{eqnarray*}

\begin{figure}[htbp]
\subfigure{\label{Figmb1}\includegraphics
  [width=0.495\textwidth]{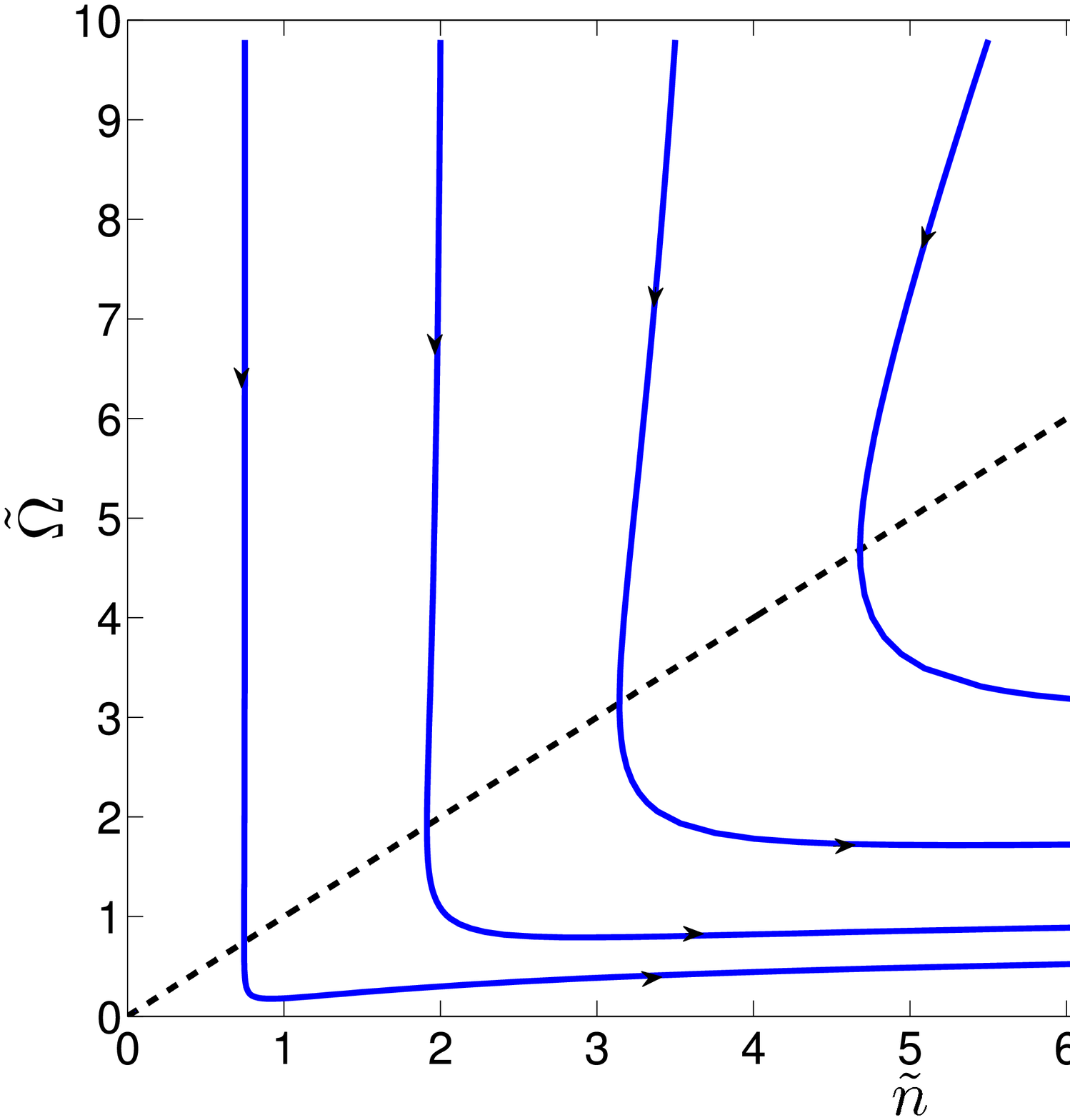}}                
\subfigure{\label{Figmb2}\includegraphics[width =
    0.495\textwidth]{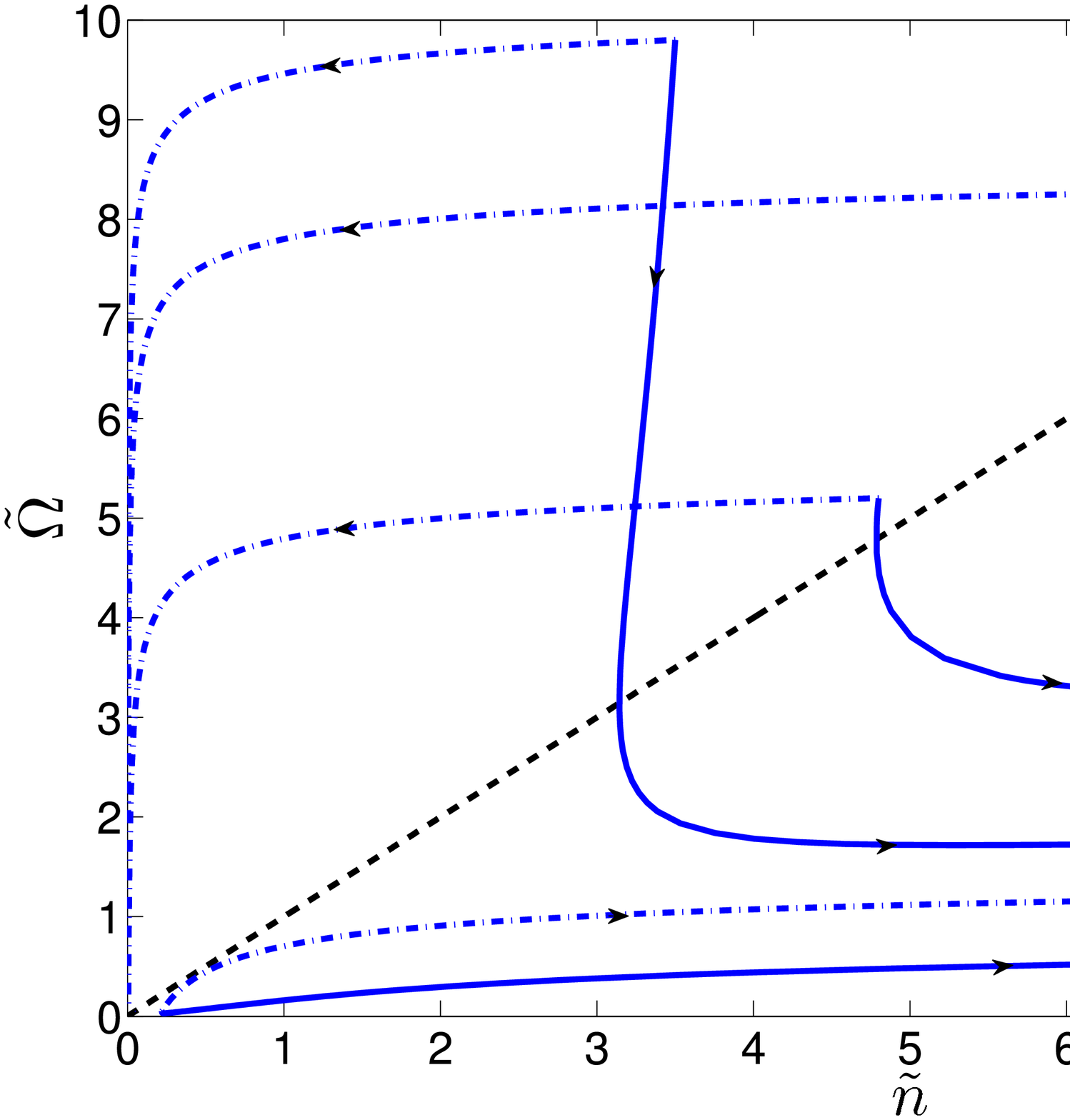}}
\caption{($\tilde \Omega,\tilde n$)-plane with $A = 100$
  for a HJ orbiting a Sun-like star. The dashed line in each
  plot corresponds to corotation ($\tilde \Omega =\tilde n$). 
  Left: magnetic braking spins the star down so that the planet finds
  itself inside corotation, where the sign of the tidal torque
  changes, and the planet is subject to tidally induced orbital decay.
  Right: solutions with and without
  magnetic braking for the same initial
  conditions. Dot-dashed lines, which
  are also curves of constant total angular momentum, are
  without braking ($A=0$) and solid lines are with braking ($A = 100$). This shows that the
  inclusion of magnetic braking is extremely important in
  determining the secular evolution of the system, and its absence
  results in a very different evolutionary history unless $\tilde
  \Omega \ll \tilde n$ in the initial state.}
 \label{fig:withMBplots}
\end{figure}

There is only one parameter ($A$) that completely characterises
the solution in the $(\tilde \Omega,\tilde n)$-plane, and its value
may be estimated as $A \simeq 100$ for a Jupiter-mass planet orbiting a Sun-like star undergoing
magnetic braking (with standard $\alpha_{mb}$ and with $Q^{\prime} = 10^{6}$).
The size of this term shows that in general magnetic braking dominates
the stellar spin evolution. We plot some solutions on the $(\tilde \Omega,\tilde
n)$-plane, restricting ourselves to $0 \leq \tilde \Omega \leq 10, 0 \leq \tilde n \leq
10$. This represents the full
range of orbits of the HJs, since $\tilde n \simeq 10$
corresponds to an orbital semi-major axis of $a \simeq 0.01$
AU, and $\tilde n \simeq 0.1$ corresponds to $a \simeq 0.2$
AU.

With the inclusion of magnetic braking (see Fig.~\ref{fig:withMBplots}), an initially rapidly
rotating solar-type star hosting a HJ for which $\tilde \Omega \geq
\tilde n$, will spin down as a result of the magnetic torque. During this stage
of spin-down the spin frequency of the star
may temporarily equal the orbital frequency of its close-in planet, but the rate of angular momentum
loss through magnetic braking will exceed the tidal rate of transfer of
angular momentum from orbit to spin. The stellar spin
continues to drop well below synchronism until the efficiency of tidal
transfer of angular momentum from orbit to spin can compensate
or overcompensate for the braking. The planet continues to
spiral into the star once it moves inside corotation, and $\tilde \Omega
\simeq const$, unless the planet has sufficient angular momentum to be
able to appreciably spin up the star. Note
that any bound orbit will eventually decay in a finite time since the
system has no stable equilibrium.

Without magnetic braking, the evolution is qualitatively different
(see right plot 
Fig.~\ref{fig:withMBplots}), with orbits initially outside 
corotation ($\tilde \Omega>\tilde n$) asymptotically approaching a stable equilibrium $\tilde \Omega = \tilde
n$ for $\tilde n \leq 3^{-\frac{3}{4}}$. Orbits initially inside
corotation with $\frac{d\tilde \Omega}{d\tilde t} >
\frac{d\tilde n}{d\tilde t} > 0$ near corotation, also approach a
stable equilibrium (though no such curves are plotted in
Fig.~\ref{fig:withMBplots}, since they occur only in the far bottom
left of the plot, near the origin). This is when the corotation radius moves inwards faster than the orbit shrinks
due to tidal friction, resulting in a final stable equilibrium
state for the system if the corotation radius ``catches up'' with the
planet. For orbits inside corotation for which this condition
is not satisfied, the planet will spiral into the star under the influence of
tides.
\vspace{-0.4cm}
\section{Tidal evolution timescales}

It is common practice to interpret the effects of tidal evolution in terms of simple
timescale estimates. The idea behind these is that if the rate of change of a
quantity $X$ is exponential, then $\dot X/X$ will be a constant, so we
can define a timescale for the evolution of $X$, given by $\tau_{X} =
X/\dot X$. If $\dot X/X \ne$ const, then $\tau_{X}$ may not accurately
respresent the evolution. A tidal inspiral time can be calculated from $\tau_{a} \equiv
-\frac{2}{13}\frac{a}{\dot a}$ (see \citealt{AJB2009} for the full evolution
equations from which these timescales are derived), since $\frac{\dot a}{a}\sim
a^{\frac{-13}{2}}$. This assumes $\Omega$ is constant, which we have
seen is unreasonable unless $\Omega \ll n$, due to magnetic
braking.

If the orbital and stellar equatorial planes are misaligned by a small
angle $i$, then dissipation of the stellar tide would align them after
a time
\small
\begin{eqnarray*}
  \tau_{i} &\equiv& -\frac{i}{\frac{di}{dt}} 
  \simeq 35.0 \: \mbox{Myr} \: 
  \left(\frac{Q^{\prime}}{10^{6}}\right)
  \left(\frac{m_{\star}}{M_{\odot}}\right)
  \left(\frac{M_{J}}{m_{p}}\right)^{2}
  \left(\frac{R_{\odot}}{R_{\star}}\right)^{3} \left(\frac{P}{1d}\right)^{4} \left(\frac{\Omega}{\Omega_{0}}\right)
  \left[1-
    \frac{P}{2P_{\star}}\left(1 - \frac{1}{\alpha}
    \right)\right]^{-1}
\end{eqnarray*} \normalsize where $\Omega_{0} = 5.8\times 10^{-6}\mbox{s}^{-1}$, and
$P,P_{\star}$ are the orbital and stellar rotational
periods. $\alpha$ is the ratio of orbit to spin angular momentum.
The validity of these timescales to accurately represent tidal
evolution is an important subject of study, since they are commonly applied to
observed systems. \cite{Jackson2008} recently found it
essential to consider the coupled evolution of $e$ and $a$ to
accurately model tidal evolution, and that both the stellar and
planetary tides must be included. They showed that the actual change
of $e$ over time can be quite different from simple
circularisation timescale considerations, due to the coupled evolution
of $a$. In the following we consider the validity of $\tau_{i}$ to accurately model tidal evolution of $i$,
using XO-3 as an illustrative example.
\vspace{-0.4cm}
\section{Application to the misaligned spin and orbit of XO-3 b}

The only system currently observed with a spin-orbit misalignment is
XO-3 \citep{Hebrard2008}, which has a sky-projected spin-orbit misalignment of
$\lambda \simeq 70^{\circ} \pm 15^{\circ}$. This system has a
very massive $m_{p} = 12.5 M_{J}$ planet on a moderately
eccentric $e = 0.287$, $P = 3.2$ d orbit around an F-type star of
mass $m_{\star} = 1.3 M_{\odot}$. Its age is
estimated to be $\tau_{\star} \simeq (2.4-3.1)$ Gyr. Note that
even if the star is rotating near breakup velocity ($P_{\star}\sim 1$ d), the planet is
still subject to tidal inspiral, since $P_{\star} > P \cos i$ (where we
henceforth assume $i = \lambda$, which may slightly \textit{underestimate} $i$).

\begin{figure}[!t]
  \begin{center}
    \subfigure[semi-major axis]{\label{FigXO-3-1}\includegraphics
      [width=0.31\textwidth,angle=-90]{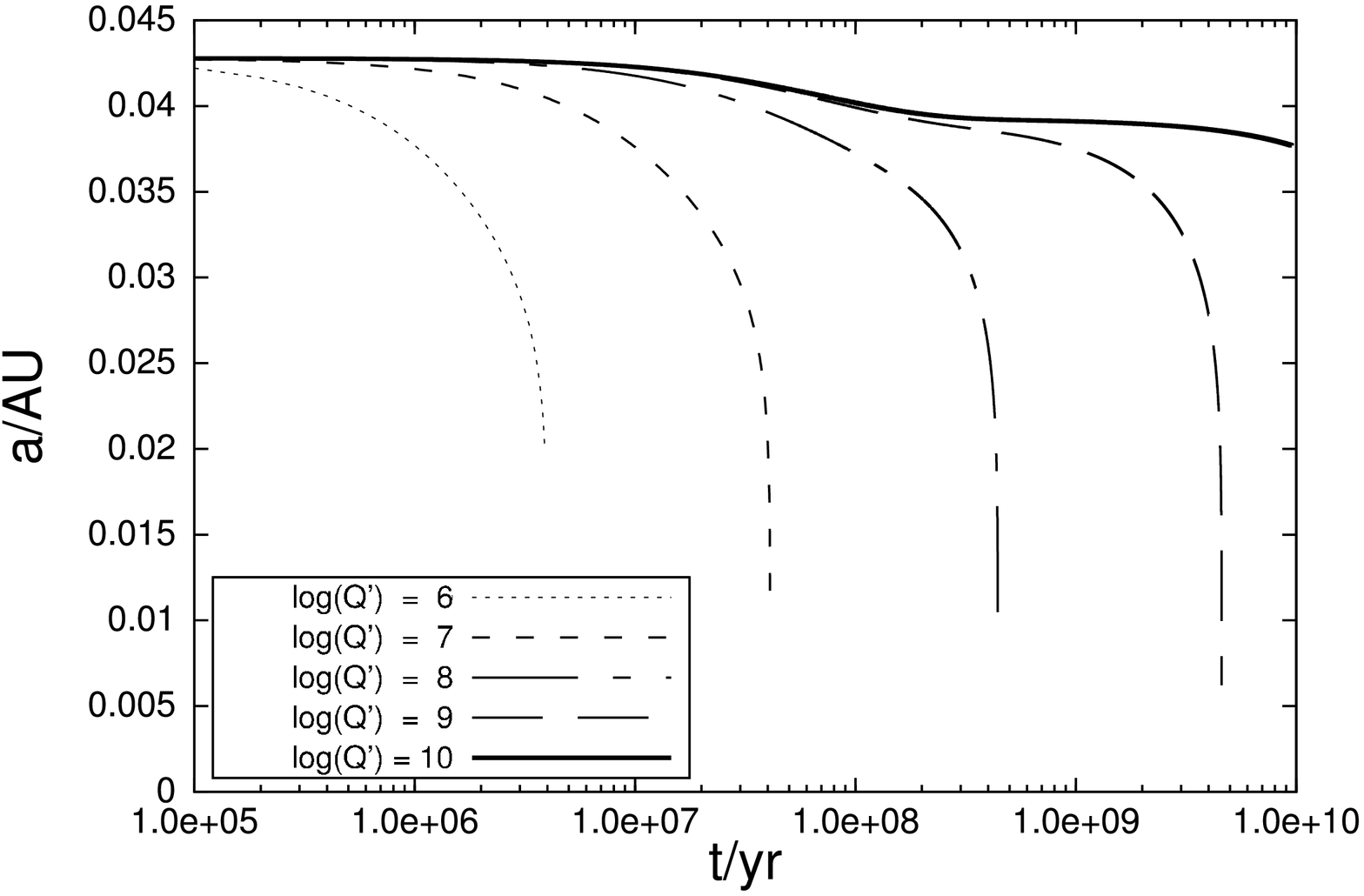}} 
    \subfigure[cosine of inclination]{\label{FigXO-3-2}\includegraphics
      [width=0.31\textwidth,angle=-90]{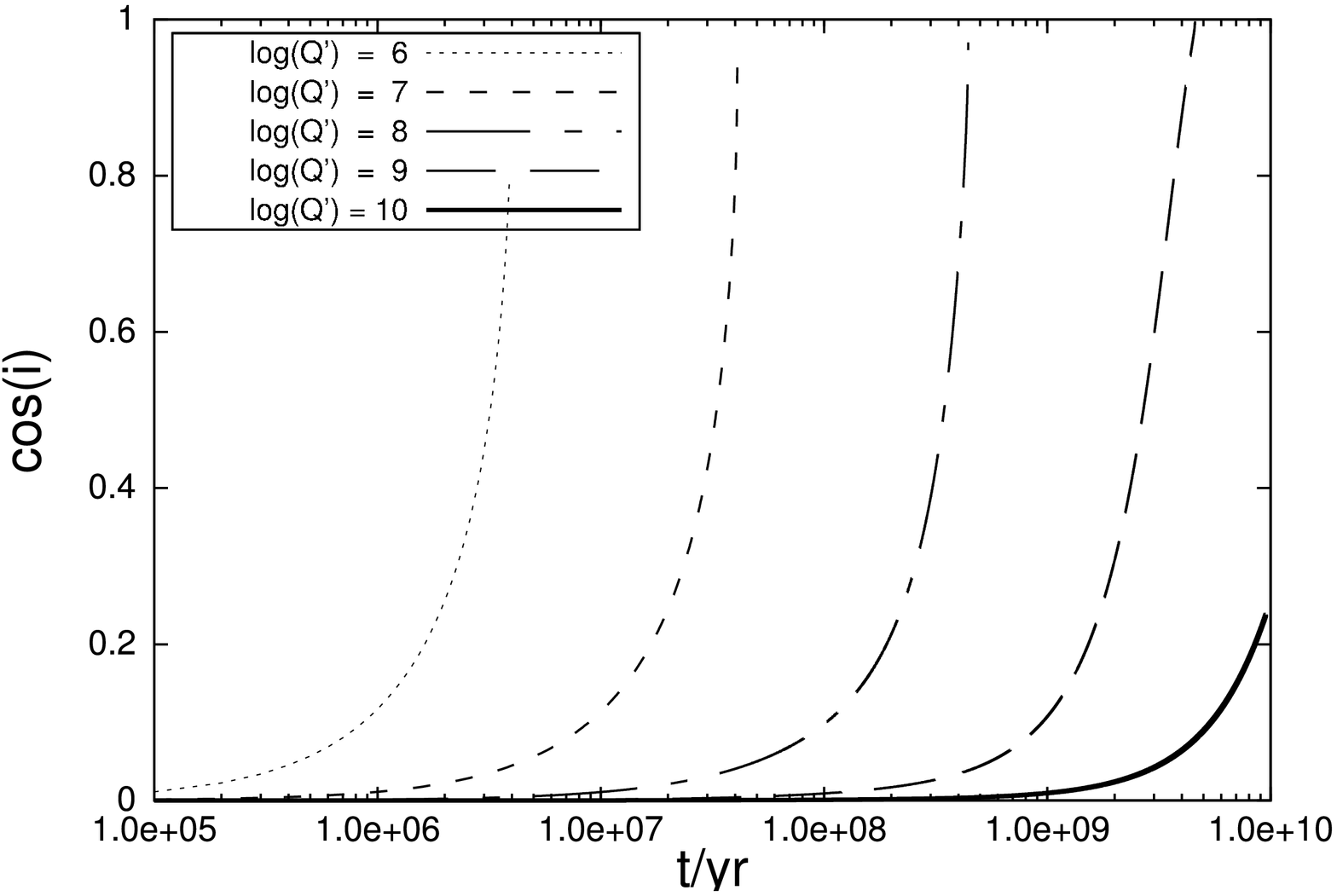}}
  \end{center}
  \caption{Tidal evolution of XO-3 b taking current values for the
    orbital properties of the system, except that $\cos i =
    90^{\circ}$ (not unreasonable since this corresponds to the upper
    limit on $\lambda$, which in any case gives a lower bound
    on $i$). Magnetic braking is included,
    and $\Omega/n = 2$ initially (results do not
    depend strongly on this choice). From (a) and (b) we
    require $Q^{\prime} \geq 10^{10}$ for the planet to survive for
    several Gyr, and maintain its high inclination. Tidal dissipation
    in star must therefore be weak to explain the current
    configuration of the system.}
  \label{FigXO-3}
\end{figure}

\cite{Hebrard2008} quote a spin-orbit alignment timescale of $\sim 10^{12}$
yr for this system, but we find that this is in error by $\sim
10^{5}$. We believe that the reason for this discrepancy is that 
their estimate was based on
assuming that the spin-orbit alignment time for XO-3 b is the same as for 
HD17156 b, which is a less massive planet on a much
wider orbit. We find $\tau_{i}\sim 15$ Myr assuming $Q^{\prime}=10^{6}$ to
align the whole star with the orbit. Integrations for this system are
given in Fig.~\ref{FigXO-3} for a variety of
stellar $Q^{\prime}$ values. For the system to survive and remain with its
current inclination for $\sim 3$ Gyr we require
$Q^{\prime}\geq 10^{10}$. These integrations highlight the
importance of considering coupled evolution
of the orbital and rotational elements, since timescales for tidal
evolution are quite different from simple estimates. Indeed, the 
actual spin-orbit alignment time from integrating the
coupled equations is about an order of magnitude smaller than that
from the simple decay estimate, due to coupled $a$ evolution.

For this system, and for typical HJs, we find that $\tau_{i} \sim
\tau_{a}$. This means that if we observe a planet, then its
survival implies that tides are unlikely to have aligned its
orbit. In addition, for systems on an accelerating inspiral
into the star, the rate of inclination evolution will have been much lower in the past; therefore,
if we observe a planet well inside corotation with a
roughly coplanar orbit, we can assume that it must have started off
similarly coplanar.

An explanation for the survival and remnant orbital inclination of
XO-3 b could be the result of inefficient tidal dissipation in
the host star. A calculation of $Q^{\prime}$ using the numerical
method of \cite{Gio2007} and a stellar model appropriate for this star
was performed \citep{AJB2009}. This predicts that the dissipation is weak, and
$Q^{\prime} \geq 10^{10}$ for most tidal frequencies for the host star
XO-3. This can explain the survival and remnant inclination of this
planet, since both inspiral and spin-orbit alignment take longer than the
current age of the system.

\vspace{-0.42cm}
\section{Conclusions}

Magnetic braking is important for calculating the long-term 
tidal evolution of HJs unless $\Omega \ll n$, and changes the qualitative
behaviour of the evolution significantly. Tidal evolution can be much faster than simple timescale
estimates predict when coupled integration of the orbital and rotational
elements is considered. In addition, we find that
$\tau_{i} \sim \tau_{a}$ for typical HJs, so the orbits of most close-in
planets have probably not aligned, and are likely to be a relic
of the migration process. This means that RM effect observations of transiting planets can potentially
distinguish between planet-planet scattering, Kozai migration and gas-disc
migration.
\vspace{-0.3cm}
\begin{acknowledgments}
A.J.B would like to thank STFC for a research studentship.
\end{acknowledgments}


\vspace{-0.7cm}
\begin{thebibliography}{}

\bibitem[Barker \& Ogilvie(2009)]{AJB2009}
  {{Barker}, A.~J., {Ogilvie}, G.~I.} 2009,
  \textit{MNRAS}, in preparation

\bibitem[Barnes(2003)]{Barnes2003}
  {{Barnes}, S.~A.} 2003,
  \textit{ApJ}, 586, 464

\bibitem[Chatterjee et al.(2008)]{CFMR2008} 
  {{Chatterjee}, S. et al.} 2008,
  \textit{ApJ}, 686, 580
  
\bibitem[Cresswell et al.(2007)]{Cresswell2007}
  {{Cresswell}, P. et al.} 2007,
  \textit{A\&A}, 473, 329

\bibitem[Dobbs-Dixon et al.(2004)]{DDLM2002}
  {{Dobbs-Dixon}, I., {Lin}, D.~N.~C., {Mardling}, R.~A.} 2004,
  \textit{ApJ}, 610, 464

\bibitem[Eggleton et al.(1998)]{Eggleton1998}
  {{Eggleton}, P.~P. and {Kiseleva}, L.~G. and {Hut}, P.} 1998,
  \textit{ApJ}, 499, 853

\bibitem[Fabrycky \& Tremaine(2007)]{Fabrycky2007} 
  {{Fabrycky}, D. and {Tremaine}, S.}, 2007,
  \textit{ApJ}, 669, 1298

\bibitem[Hebrard et al.(2008)]{Hebrard2008}
  {{Hebrard}, G. et al.} 2008, 
  \textit{A\&A}, 488, 763

\bibitem[Hut(1980)]{Hut1980}
  {{Hut}, P.} 1980, 
  \textit{A\&A}, 92, 167
  
\bibitem[Hut(1981)]{Hut1981}
  {{Hut}, P.} 1981, 
  \textit{A\&A}, 99, 126

\bibitem[Jackson et al.(2008)]{Jackson2008}
  {{Jackson}, B., {Greenberg}, R., {Barnes}, R.} 2008, 
  \textit{ApJ}, 678, 1396

\bibitem[Juric \& Tremaine(2008)]{JT2008}
  {{Juri{\'c}}, M., {Tremaine}, S.} 2008,
  \textit{ApJ}, 686, 603

\bibitem[Lin, Bodenheimer \& Richardson(1996)]{LBR1996}
  {{Lin}, D.~N.~C., {Bodenheimer}, P., {Richardson}, D.~C.
  } 1996, 
  \textit{Nature}, 380, 606

\bibitem[Lubow \& Ogilvie(2001)]{LubOg2001}
  {{Lubow}, S.~H. and {Ogilvie}, G.~I.} 2001,
  \textit{ApJ}, 560, 997

\bibitem[Mardling \& Lin(2002)]{ML2002}
  {Mardling, R.~A., Lin, D.~N.~C.} 2002,
  \textit{ApJ}, 573, 829

\bibitem[Mayor \& Queloz(1995)]{MQ1995}
  {{Mayor}, M., {Queloz}, D.} 1995, 
  \textit{Nature}, 378, 355

\bibitem[Ogilvie \& Lin(2007)]{Gio2007}
  {{Ogilvie}, G.~I., {Lin}, D.~N.~C.} 2007, 
  \textit{ApJ}, 661, 1180
  
\bibitem[Ogilvie \& Lin(2004)]{Gio2004}
  {{Ogilvie}, G.~I., {Lin}, D.~N.~C.} 2004, 
  \textit{ApJ}, 610, 477
  
\bibitem[Papaloizou et al.(2008)]{PapProt2007} 
 {{Papaloizou}, J.~C.~B. et al.} 2007,
  \textit{Protostars and Planets V}, 655

\bibitem[Skumanich(1972)]{Skumanich1972}
  {{Skumanich}, A.} 1972, 
  \textit{ApJ}, 171,565

\bibitem[Verbunt \& Zwaan(1981)]{VerbuntZwaan1981}
  {{Verbunt}, F., {Zwaan}, C.} 1981,
  \textit{A\&A}, 100, L7

\bibitem[Weber \& Davis(1967)]{WD1967}
  {{Weber}, E.~J., {Davis}, L.~J.} 1967, 
  \textit{ApJ}, 148, 217

\bibitem[Winn(2008)]{Winn2008}
  {{Winn}, J.~N.} 2008, 
  ArXiv e-prints, 0807.4929

\bibitem[Zahn(2008)]{Zahn2008}
  {{Zahn}, J.-P.} 2008, 
  \textit{Tidal dissipation in binary systems}, EAS Publications Series, 29, 67
\end{thebibliography}
\end{document}